\documentclass[
 reprint, amsmath,amssymb, aps,
]{revtex4-1}

\usepackage[colorlinks=true,linkcolor=black, citecolor=black,
urlcolor=black]{hyperref}

\usepackage{multirow,graphics}
    \newcommand{\beq}{\begin{equation}}
    \newcommand{\eeq}{\end{equation}}
    \newcommand\beqa{\begin{eqnarray}}
    \newcommand\eeqa{\end{eqnarray}}

\newcommand{\la}[1]{\label{#1}}
\newcommand{\eq}[1]{(\ref{#1})}

\newcommand{\sym}{$\mathcal{N}=4$ SYM}

\usepackage{amstext}
\usepackage{amssymb}
\usepackage{amsmath}
\usepackage{graphicx}

\usepackage[usenames,dvipsnames]{color}

\begin{document}
\makeatletter
     \@ifundefined{usebibtex}{\newcommand{\ifbibtexelse}[2]{#2}} {\newcommand{\ifbibtexelse}[2]{#1}}
\makeatother

\preprint{Imperial/TP/13/SL/02}

\newcommand{\footnotea}[1]{\ifbibtexelse{\footnote{#1}}{
\newcommand{\textfootnotea}{#1}
\cite{thefootnotea}}}
\newcommand{\footnoteb}[1]{\ifbibtexelse{\footnote{#1}}{
\newcommand{\textfootnoteb}{#1}
\cite{thefootnoteb}}}
\newcommand{\footnotebis}{\ifbibtexelse{\footnotemark[\value{footnote}]}{
\cite{thefootnoteb}}}

\def\e{\epsilon}
     \def\bT{{\bf T}}
    \def\bQ{{\bf Q}}
    \def\wT{{\mathbb{T}}}
    \def\wQ{{\mathbb{Q}}}
    \def\ttQ{{\bar Q}}
    \def\tQ{{\tilde \bP}}
    \def\bP{{\bf P}}
    \def\CF{{\cal F}}
    \def\cC{\CF}
     \def\Tr{\text{Tr}}
     \def\l{\lambda}
\def\hbZ{{\widehat{ Z}}}
\def\bZ{{\resizebox{0.28cm}{0.33cm}{$\hspace{0.03cm}\check {\hspace{-0.03cm}\resizebox{0.14cm}{0.18cm}{$Z$}}$}}}
\newcommand{\rb}{\right)}
\newcommand{\lb}{\left(}
\newcommand{\gT}{T}\newcommand{\gQ}{Q}

\title{NNLO BFKL Pomeron eigenvalue in N=4 SYM}

\author{ Nikolay Gromov$^{a,b}$, Fedor Levkovich-Maslyuk$^{a}$, Grigory Sizov$^{a}$,}

\affiliation{
\(^{a}\)Mathematics Department, King's College London, The Strand, London WC2R 2LS, UK
\\
\(^{b}\)
 St.Petersburg INP, Gatchina, 188300, St.Petersburg, Russia
               }

\begin{abstract}
We obtain an analytical expression for the Next-to-Next-to-Leading order of the Balitsky-Fadin-Kuraev-Lipatov (BFKL) Pomeron eigenvalue in planar \sym\;using Quantum Spectral Curve (QSC) integrability based method.
The result is verified with more than $60$ digits precision using the numerical method developed by us in a previous paper.
As a byproduct we developed a general analytic method of solving the QSC perturbatively.
\end{abstract}

 \maketitle

\section{Introduction}
QCD is notorious for being hard to explore analytically: perturbative calculations become impossibly complex after first few loop orders. However, there are regimes in which one can probe all orders of perturbation theory analytically.
The Balitsky-Fadin-Kuraev-Lipatov (BFKL) equation is applicable in processes like Deep Inelastic Scattering or hadronic dijet production, which are characterized by a presence of at least two widely separated energy scales.
The large logarithm of ratio of these energy scales
$\Delta y$
enters into   perturbative expansion, so in order to make sense of the perturbation theory one has to resum powers of $\Delta y$ in every order of perturbation theory.
Schematically these large corrections exponentiate to an expression of the form
\beq\la{fint}
 d\sigma(\Delta y, p_{\perp})\propto\sum\limits_{n=-\infty}^{+\infty}e^{i n\tilde \phi}\int\limits_{-\infty}^{\infty}d\nu\; Q_\nu(p_{\perp}) e^{4 g^2 \chi(\nu,n)\Delta y}
 \eeq
where  $g=\frac{\sqrt\lambda}{4\pi}$ with $\lambda$ being 't Hooft coupling constant.
 In particular this can be done in the case of the high-energy hadron-hadron scattering \cite{bfkl}.
In this case $Q$ absorbs the trivial dependence on the transverse momenta
and the nontrivial part $\chi$ is the so-called LO BFKL eigenvalue \cite{Jaroszewicz:1982gr,Lipatov:1985uk}
 \beq\la{LO0}
 \chi^{LO}(\nu,n)\!=\!2\psi(1)\!-\!\psi\left(\!\frac{n+1+i\nu}{2}\!\right)\!-\!\psi\left(\!\frac{n+1-i\nu}{2}\!\right).
 \eeq

In this paper we focus on the case $n=0$.
Taking into account the Next-to-Leading, Next-to-Next-to-Leading corrections we get a similar structure, where the BFKL eigenvalue $\chi$ in the exponent gets corrected. One usually introduces $j(i\nu)$, related to the BFKL eigenvalue as
 \beq\nonumber
 \frac{j(i\nu)-1}{4g^2}=\chi^{\textrm{LO}}(\nu,0)+g^2\chi^{\textrm{NLO}}(\nu,0)+g^4\chi^{\textrm{NNLO}}(\nu,0)+\dots\;.
 \eeq

 The Next-to-Leading BFKL was obtained after 9 years of laborious calculations in \cite{Fadin:1998py,Ciafaloni:1998gs,Kotikov:2000pm,moreKotikov:2002,Kotikov:2002ab}; the result in modern notation is presented below in the text \eqref{jexp}. The corrections turned out to be
 numerically rather large compared to the LO, which makes one question the validity of the whole BFKL resummation procedure and its applicability for phenomenology.


This and other indications make it clear that just NLO may not be enough to match experimental predictions. It is important to understand the general structure of BFKL expansion terms and this paper is concerned with NNLO BFKL eigenvalue in \sym ~--- a more symmetric analog of QCD.
 Notably, it was observed in \cite{Kotikov:2002ab} that the ${\cal N}=4$ SYM reproduces correctly the part of the QCD result with maximal transcendentality.
 In particular the LO coincides exactly in the two theories.

Another way of extracting the Pomeron eigenvalue, technically more convenient, is due to the observation of
\cite{Kotikov:2000pm} who reformulated the problem in terms of a certain analytical continuation
of
anomalous dimensions of twist-2 operators.
Fortunately, in planar ${\cal N}=4$ SYM the problem of computing the anomalous dimensions
is solved for finite coupling and any operator by the Quantum Spectral Curve (QSC) formalism \cite{PmuPRL,PmuLong}.

In order to obtain the BFKL eigenvalue in \sym~from the anomalous dimension of twist operators we consider the dimension $\Delta(S)$ of twist-two operator ${\cal O}=\textrm{Tr} ZD_+^SZ$.
The inverse function $S(\Delta)$ is known to approach $-1$ perturbatively for $\Delta$ in the range $[-1,1]$
and thus the map to the BFKL regime is given by $\Delta=i\nu$ and $j=2+S(\Delta)$.
Then the goal is to compute $j(\Delta)$ as a series expansion in $g^2$.
Indeed, from the QSC formalism it was shown in \cite{Alfimov:2014bwa} that one reproduces correctly the LO \eq{LO0}.
Here we use some shortcuts to the direct approach of \cite{Alfimov:2014bwa} to push the calculation
to NNLO order, which already gives useful new information about the QCD result.

An essential for us observation was made in \cite{Costa:2012cb}\footnote{We are grateful to S.~Caron-Huot for bringing our attention to this paper}
where it was pointed out that both LO and NLO results can be represented as a simple linear combination
of the nested harmonic sums. Let us stress again that in our notation $\Delta$ is the full conformal dimension of the twist-two operator, related to the anomalous dimension $\gamma$ as $\Delta=2+S+\gamma$. Then the expansion of $j(\Delta)$ can be written as
\beq \la{jexp}
j(\Delta)=1+\sum\limits_{\ell=1}^\infty g^{2\ell}\left[F_\ell\left(\frac{\Delta-1}{2}\right)+F_\ell\left(\frac{-\Delta-1}{2}\right)\right]
\eeq
with the two first known orders given by \cite{Costa:2012cb}
\beqa\la{LO}
&&F_1=-4 S_1\\
\nonumber&&\frac{F_2}{4}=-\frac{3}{2}\zeta_3+\pi^2\ln 2+\frac{\pi^2}{3}S_1+2S_3+\pi^2 S_{-1}-4 S_{-2,1}
\eeqa
where
\beq\nonumber
S_{a_1,a_2,\dots,a_n}(x)=\sum\limits_{y=1}^x\frac{(\textrm{sign}\left(a_1\right))^y}{y^{|a_1|}}S_{a_2,\dots,a_n}(y)\;\;,\;\;S(x)=1\;.
\eeq
We define harmonic sums for non-integer and negative arguments by the standard widely accepted prescription, namely analytical continuation from positive even integer values as in \cite{Kot1,ancont}. These analytically continued sums, which we denote as $S_{a_1,a_2,\dots}$, are denoted by $\bar S^+$ in \cite{Kot1}, see e.g. Eq. (21) in that paper. A compatible but more general definition is given in \cite{Blum}. This prescription for analytic continuation is also implemented in the Mathematica files attached to the present paper \cite{NB}.

We assume the NNLO order can also be written in this form. After that
we only have to fix a finite number of coefficients which we do by expanding the QSC
around some values of $\Delta$ where the result simplifies.
Then we verify our result by comparing it with extremely high precision numerical evaluation proving this assumption to be correct.

\section{Quantum Spectral Curve Generalities}
As it was already mentioned in the introduction there is a known relation between the
anomalous dimensions of the  twist-2 operators and the BFKL pomeron eigenvalue.
Here we describe the Quantum Spectral Curve (QSC) solution of the spectral problem -
a simple set of equations giving  the full spectrum of the anomalous dimensions of the theory developed in \cite{PmuPRL,PmuLong}.
Below we limit ourselves to the $sl(2)$ sector of the theory.

The simplest ingredient of the QSC is a set of $4$ functions $\bP_a$
of the spectral parameter $u$ which can be conveniently written as a convergent  series expansion
\beq\nonumber
\bP_a(u)=\sum_{n=\tilde M_a}^\infty \frac{c_{a,n}}{x^n(u)}\;\;,\;\;x(u)=\frac{u+\sqrt{u-2g}\sqrt{u+2g}}{2g}\;.
\eeq
We see that $\bP_a$ has a branch cut and is power-like at infinity. The
constants $\tilde M_a$ control the global charges of the state. For the case of the twist-2 operators,
 $\tilde M_a=\{2,1,0,-1\}$. The problem of solving the QSC consists in finding the coefficients $c_{a,n}$.
They can be fixed in the following steps \cite{Gromov:2015wca}:

First, find $4$ linear independent analytic in the upper half plane solutions of the linear finite difference equation
\beq\la{QQPPQ}
{\cal Q}_{a|i}(u+\tfrac{i}{2})-{\cal Q}_{a|i}(u-\tfrac{i}{2})=-\bP_a(u)\bP^b(u){\cal Q}_{b|i}(u+\tfrac{i}{2})
\eeq
where $i$ labels the $4$ solutions. Here and everywhere in this paper indices are raised with a $ 4 \times 4$ matrix $\chi^{ab}=(-1)^a \delta_{a,5-b}$. 
 The solutions can be always chosen to have ``pure" asymptotics, which means that with exponential precision
the large $u$ (asymptotic) expansion of ${\cal Q}_{a|i}$ has the form
\beq\la{Qailarge}
{\cal Q}_{a|i}\simeq u^{\hat M_i-\tilde M_a}\sum_{n=0}^\infty \frac{A_{a,i,n}}{u^n}\;.
\eeq
In the generic situation, it is always possible to choose the $4$ solutions of that equation
in this form. In what follows we assume this to be done. We require in addition that $\hat M_i$ encode the conformal charges of the operators i.e.
$2\hat M_i=\{\Delta -S+2,\Delta +S,-\Delta -S+2,S-\Delta \}$, where $\Delta$ is the dimension of the operator in question.
This requirement fixes some of the leading coefficients $c_{a,0}$.
There is an obvious rescaling freedom of ${\cal Q}_{a|i}$ which can be partially fixed by requiring that
\beq\la{Qainorm}
{\cal Q}_{a|i}{\cal Q}^{a|j}=-\delta^j_{\;i}\;.
\eeq

Next, one finds $4$ Q-functions by dividing either side of \eq{QQPPQ} by $\bP_a(u)$
\beq\la{bQi}
\bQ_i(u)=-\bP^b(u){\cal Q}_{b|i}(u+i/2)\;.
\eeq
They are now given implicitly in terms of the coefficients $c_{a,n}$. The main constraint
comes from the condition that the analytic continuation which we denote as $\tilde \bQ_i$
is a linear combination with periodic coefficients of $\bQ_i$ themselves
\beq\la{Qt}
\tilde\bQ_i=\omega_{ij}\bQ^j=-\tilde\bP^b(u){\cal Q}_{b|i}(u+i/2)
\eeq
where $\tilde\bP^b(u)$ is the same as $\bP^b(u)$ with $x$ replaced by $1/x$.
In particular from \eq{Qt} we have
\beqa\la{Q13}
&&\tilde \bQ_1(u)=\omega_{12} \bQ_3(u)+\omega_{14} \bQ_1(u)-\omega_{13} \bQ_2(u)\;,\\
&&\nonumber\tilde \bQ_3(u)=\omega_{34} \bQ_1(u)-\omega_{14} \bQ_3(u)+\omega_{13} \bQ_4(u)\;.
\eeqa
As we will see, $\omega_{ij}$ can be eliminated. To show this we will need only to know that $\omega_{ij}$
is $i$-periodic, anti-symmetric and should satisfy
\beqa\la{omegaconstr}
\omega_{ij}\omega^{jk}=\delta_i^k\;\;,\;\;\omega_{23}=\omega_{14}\;.
\eeqa
For all physical operators $\omega_{ij}$ should go to a constant at large $u$, however
it was emphasized in particular in \cite{Gromov:2015wca} and \cite{Gromov:2014bva} that for non-integer $S$
one should allow for an exponential growth of $\omega_{24}$
as otherwise the system has no solution.
Note that \eq{omegaconstr} implies that $\omega^{24}=\omega_{13}$ should decay exponentially at infinity.
It is also known that $\omega_{14}$ decays at infinity \cite{Gromov:2015wca}.
Condition \eq{Qt} in fact imposes infinitely many constraints on the coefficients $c_{a,n}$
fixing them completely as well as the function $\Delta(S)$ or $S(\Delta)$.


\section{Analytical Data from QSC}

We describe now the details of our analytical method.
We will focus on some particular points $\Delta_0=1,3,5,7$. It can be seen already
from the LO \eq{LO0} that the function $S(\Delta)$ is singular at these points, however the coefficients
of the expansion are relatively simple and are given by $\zeta$-functions.
We will perform a double expansion first in $g$ up to the order $g^6$
and then in $\delta=\Delta-\Delta_0$.

\paragraph{General iterative procedure for solving QSC.}
We describe a procedure which for some given $\bP_a$ (or equivalently $c_{a,n}$)
takes as an input some approximate solution of \eq{QQPPQ} ${\cal Q}_{a|i}^{(0)}$ valid up to the order $\epsilon^n$ (where $\epsilon$ is some small expansion parameter) and produces as an output new ${\cal Q}_{a|i}^{}$
accurate to the order $\epsilon^{2n}$. The method is very general and in particular is suitable for perturbative
expansion around any background.

Let $dS$ be the mismatch in the equation \eq{QQPPQ}, i.e.
\beq \la{eqdS}
{\cal Q}_{a|i}^{(0)}(u+\tfrac{i}{2}) -
{\cal Q}_{a|i}^{(0)}(u-\tfrac{i}{2})
+\bP_a\bP^b {\cal Q}_{b|i}^{(0)}(u+\tfrac{i}{2})=dS_{a|i},
\eeq
where $dS_{a|i}$ is small $\sim \epsilon^n$. We can always represent the exact solution in the form
\beq\la{Qc}
{\cal Q}_{a|i}(u)={\cal Q}_{a|i}^{(0)}(u)+{b_i^{\;j}}(u+\tfrac{i}{2})\;{\cal Q}_{a|j}^{(0)}(u)
\eeq
where the unknown functions $b_i^{\;j}$ are also small.
After plugging this ansatz into the equation \eqref{eqdS} we get
\beq
\left(b_i^{\;j}(u)-b_i^{\;j}(u+i)\right){\cal Q}_{a|j}^{+(0)}=dS_{a|i}+dS_{a|j}b_i^{\;j}\;.
\eeq
Since $b_i^{\;j}$ is small it can be neglected in the r.h.s. where it multiplies another small quantity.
Finally multiplying the equation by ${\cal Q}^{(0)a|k}$ and using \eq{Qainorm} we arrive at
\beq\nonumber
{b_i^k}(u+i)-b_i^k(u)=-dS_{a|i}(u){{\cal Q}^{(0){a|k}}}\left(u+\tfrac{i}{2}\right)+{\cal O}(\epsilon^{2n})\;.
\eeq
We see that the r.h.s. contains only the known functions $dS$ and ${\cal Q}^{(0)}$ and does not contain
$b$ which means that the original $4$th order finite difference equation is reduced to a set of independent $1$st
order equations! In most interesting cases the first order equation can be easily solved.
After ${\cal Q}_{a|i}$ is found one can use \eq{bQi} to find $\bQ_i$.

\paragraph{Iterations at weak coupling.}
For our particular problem we will take either $\epsilon=g$ or $\epsilon=\delta$.
Applying this procedure a few times we generate $\bQ_i$
for sufficiently high order both in $g$ and in $\delta$.
Finally, by ``gluing'' $\bQ_i$ and $\tilde\bQ_i$ on the cut
we find $c_{a,n}$ and $S(\Delta)$ also as a double expansion.

For the above procedure we need the leading order ${\cal Q}^{(0)}_{a,i}$.
One can expect that to the leading order in $g$ the solution should be very simple -
indeed the branch cuts collapse to a point making most of the functions polynomial or
having very simple singular structure. Also one can use that to the leading order in $g$ functions
$\bP_a$ are very simple and are already known from \cite{Alfimov:2014bwa} for any $\Delta$.
By making a simple ansatz for $\bQ_i$ we found for $\Delta_0=1$ to the leading order
\beq
\bQ_1\simeq u,\;\bQ_2\simeq 1/u,\;\bQ_3\simeq 1,\;\bQ_4\simeq 1/u^2\;.
\eeq
For $\Delta_0=3,5,\dots$ the solution involves also the $\eta$-functions
introduced in the QSC context in \cite{Leurent:2013mr,Marboe:2014gma}
\beq
\eta_{s_1,\dots,s_k}(u)=\sum_{n_1>n_2\dots n_k\ge 0}\frac{1}{(u+i n_1)^{s_1}\dots (u+i n_k)^{s_k}}.
\eeq
which are related in a simple way to the nested harmonic sums. For $\Delta=3$ we found
\beqa
&&\bQ_1\simeq u^2,\;\bQ_2\simeq u^2 \eta_{1,3}-i-\frac{1}{2 u},\\
\nonumber &&\bQ_3\simeq  u^2 \eta_{1,2}-i u-\frac{1}{2},\;\bQ_4\simeq u^2 \eta_{1,4}-\frac{i}{u}-\frac{1}{2 u^2}\;,
\eeqa
which reflects the general structure of the expansion of $\bQ_i$ around integer $\Delta$'s
which contain only $\eta_{1,2},\;\eta_{1,3}$ and $\eta_{1,4}$ with polynomial coefficients.
As it was explained in \cite{Leurent:2013mr,Marboe:2014gma} the $\eta$-functions are closed under all essential for us operations:
the product of any two $\eta$-functions can be written as a sum of $\eta$-functions,
and most importantly one can easily solve equations of the type
\beq
f(u+i)-f(u)=u^n \eta_{s_1,\dots,s_k}
\eeq
for any integer $n$ again in terms of a sum of powers of $u$ multiplying $\eta$-functions (which we call
$\eta$-polynomials). For example
for $n=-1$ and $k=1,\;s_1=1$ we get $f=-\eta_2-\eta_{1,1}$ etc.
Thus for these starting points
we are guaranteed to get $\eta$-polynomials on each step of the general
procedure described above.

Proceeding in this way we computed $\bQ_i$
up to the order $g^6$ and $\delta^{10}$ for $\Delta=3,5,7$.
After that we fix the coefficients in the ansatz for $\bP_a$ from analyticity requirements
described below.

\paragraph{Fixing remaining freedom.}
Here we will describe how to use $\bQ_i$ found before to finally extract relation between $S$ and $\Delta$
and the constants $c_{a,n}$. This is done by using a relation between $\bQ_i$ and their analytical continuations $\tilde \bQ_i$.
On the one hand we have the relation \eq{Q13}.
On the other hand we can use the $u\to-u$
symmetry\footnote{more generally one can also use complex conjugation symmetry} of the twist-2 operators to notice that
$\bQ_i(-u)$ should satisfy the same finite difference equation as $\bQ_i(u)$
and thus we should have $\bQ_i(u)=\Omega_i^{j}(u)\bQ_j(-u)$
where $\Omega_i^j(u)$ is a set of periodic coefficients.
As $\bQ_i(u)$ has a power-like behavior at infinity, $\Omega_{i}^j(u)$
should not grow faster than a constant.
Furthermore, since $\bQ_i$ has a definite asymptotic \eq{Qailarge}
only diagonal elements of $\Omega_{i}^i(u)$ can be nonzero at infinity.
 Combining these relations we find
\beqa
\tilde \bQ_A(u)=\alpha_{A}^i \bQ_i(-u)\;\;,\;\;A=1,3\;,
\eeqa
where $\alpha_{A}^j=\omega_{Ai}\chi^{ik}\Omega_k^{j}$ are $i$-periodic (as a combination of $i$-periodic functions),
analytic (as both $\tilde \bQ_a(u)$ and $\bQ_a(-u)$ should be analytic in the lower-half-plane) and
growing not faster than a constant at infinity which implies that they are
constants. Furthermore most of them are zero because only $\omega_{12},\;\omega_{34}$
and $\Omega^i_i$ are non-zero at infinity. Thus we simply get
\beqa\la{bQ13}
\tilde \bQ_1(u)=\alpha_{13} \bQ_3(-u)\;\;,\;\;
\tilde \bQ_3(u)=\alpha_{31} \bQ_1(-u)\;.
\eeqa
Next we note that if we analytically continue this relation and change $u\to-u$
we should get an inverse transformation which
implies $\alpha_{13}=1/\alpha_{31}\equiv \alpha$.
The coefficient $\alpha$ depends on relative normalization of $\bQ_1$ and  $\bQ_3$.
Let us see how to use the identity \eq{bQ13} to constrain the constants $c_{a,i}$.
We observed that all the constants are fixed from the requirement of regularity at the origin
of the combinations $\bQ_1+\tilde \bQ_1$ and $\frac{\bQ_1-\tilde \bQ_1}{\sqrt{u^2-4g^2}}$,
which now can be written as
\beq\nonumber
\bQ_1(u)+\alpha \bQ_3(-u)={\rm reg}\;\;,\;\;
\frac{\bQ_1(u)-\alpha \bQ_3(-u)}{\sqrt{u^2-4g^2}}={\rm reg}\;.
\eeq
This relation is used in the following way: one first expands in $g$ the l.h.s.
and then in $u$ around the origin. Then requiring the absence of the negative
powers will fix $\alpha$, all the coefficients $c_{a,n}$,
and the function $\Delta(S)$! So we can completely ignore $\omega_{ij}$, $\bQ_2$,
and $\bQ_4$ in this calculation. This observation can be used in more general situations
and allows avoiding construction of $\omega_{ij}$, and in particular can simplify
the numerical algorithm of \cite{Gromov:2015wca} considerably.

\begin{widetext}
\paragraph{Constraints from poles.}
We use the procedure described above
 to compute the expansion of $S(\Delta)$ around $\Delta_0=3,5,7$. In particular
for $\Delta=5+\epsilon$ we computed the first $8$ terms
\beqa\la{D5exp}
&&\chi^{\rm NNLO}=-\frac{1024}{\epsilon ^5}+\frac{64 \left(4 \pi
   ^2-33\right)}{3 \epsilon ^3}+\frac{16 \left(-36
   \zeta _3+2 \pi ^2+31\right)}{\epsilon
   ^2}+\frac{-288 \zeta _3+\frac{232 \pi ^4}{45}-16
   \pi ^2-296}{\epsilon }\\
\nonumber&&-\frac{2}{15} \left[20
   \left(4 \pi ^2-75\right) \zeta _3+6300 \zeta
   _5+\pi ^4-215 \pi ^2+285\right]
   +\dots\;. 
\eeqa
The terms with $\epsilon,\;\epsilon^2,$ and $\epsilon^3$
which we also evaluated explicitly
are omitted for the sake of brevity.
We also reproduced expansions extracted from \cite{Marboe:2014sya}
for  $\Delta=1$. In our calculations we used several Mathematica packages for manipulating harmonic sums and multiple zeta values \cite{packages}.
\end{widetext}
\section{The result}
By observing \eqref{LO} for LO and NLO we notice that the transcendentality of these expressions is uniform if one assigns to $S_{a_1,\dots,a_k}$ transcendentality equal to $\sum\limits_{j=1}^k |a_j|$.
The principal assumption of our calculation states that $F_3(x)$ can also be written
as a linear combination of nested harmonic sums with coefficients made out of several transcendental constants
$
\pi^2,\log(2),\zeta_3,\zeta_5,\textrm{Li}_{4}\!\left(\frac{1}{2}\right),\textrm{Li}_5\!\left(\frac{1}{2}\right)
$
 of uniform transcendentality $5$.
 The final basis obtained after taking into account the constants contains $288$ elements.

Hence we build the linear combination  of these basis elements with free coefficients
and constrained them by imposing the expansion at $\Delta=1,3,5,7$ to match the results of the analytic expansion of QSC (in particular, requiring \eq{D5exp}). This gave an overdefined system of linear equations for the unknown coefficients which
happen to have a unique solution presented below:
\beqa
\nonumber&&\frac{F_3(x)}{256}=-\frac{5
   S_{-5}}{8}-\frac{S_{-4,1}}{2}+\frac{ S_1
   S_{-3,1}}{2}+\frac{S_{-3,2}}{2}-\frac{5S_2
   S_{-2,1}}{4} \\
\nonumber&&+\frac{
   S_{-4} S_1}{4}+\frac{S_{-3}
   S_2}{8} +\frac{3
   S_{3,-2}}{4}-\frac{3S_{-3,1,1}}{2} -S_1
   S_{-2,1,1}\\
\la{res}&&+S_{2,-2,1}+3 S_{-2,1,1,1}-\frac{3S_{-2}
   S_3}{4} -\frac{S_5}{8}+\frac{ S_{-2} S_1 S_2}{4}\\
\nonumber&&+\pi ^2\left[\frac{S_{-2,1}}{8}  -\frac{7S_{-3}}{48}
   -\frac{S_{-2}S_1}{12}
   +\frac{S_1 S_2}{48}  \right]\!-\!\pi ^4 \left[\frac{2S_{-1}}{45} -\frac{  S_1}{96}\right]\\
\nonumber&&+\zeta _3\left[-\frac{7S_{-1,1}}{4}
   +\frac{7S_{-2}}{8}
   +\frac{7S_{-1}
   S_1}{4}  -\frac{S_2}{16}\right]\\
\nonumber&&+\left[2
   \text{Li}_4\!\!\left(\tfrac{1}{2}\right)
   -\frac{ \pi
   ^2  \log ^2\!2}{12}+\frac{\log
   ^4\!2}{12}
\right]
   \left(S_{-1}- S_1\right)\\
\nonumber&&+\frac{\log ^5\!2}{60}-\frac{\pi ^2
   \log ^3\!2}{36} -\frac{2 \pi ^4 \log 2}{45}
   -\frac{\pi ^2
   \zeta _3}{24}+\frac{49 \zeta _5}{32}-2
   \text{Li}_5\!\!\left(\tfrac{1}{2}\right)\;.
\eeqa
The simplicity of the final result is quite astonishing: only $37$ coefficients out of $288$
 turned out to be nonzero. Furthermore, they are significantly simpler than the coefficients appearing in the series expansion around the poles
 \eq{D5exp}. These are all clear and expected indications of the correct result similar to what was observed in the usual perturbation theory
\cite{Kotikov:2007cy}. In addition we also performed the numerical test described below.

\section{Numerical tests}
Using the method of \cite{Gromov:2015wca} we evaluated $40$ values of spin $S$
for various values of the coupling $g$ in the range $(0.01,0.025)$ with exceptionally high $80$
digits precision and then fit this data to get the following prediction for the N$^n$LO BFKL coefficients
at the fixed value of $\Delta=0.45$:
\beq
\begin{array}{l|l|l}

& \rm value & \rm error\\ \hline
{\rm N^2LO} &
\begin{array}{r}
10774.6358188471766379575931271924\;\;
\\56995929170948057653783424533229
\end{array}
 & 10^{-61}
\\ \hline
{\rm N^3LO} &
\begin{array}{r}
-366393.20520539170389379035074785\;\;
\\44549935531959333919163403836
\end{array}
 & 10^{-56}
\\ \hline
{\rm N^4LO} &
\begin{array}{r}
1.33273635568112691569404431036982\;\;
\\8561521940588979476878854\times
   10^7
\end{array}
 & 10^{-51}
\\ \hline
{\rm N^5LO} &
\begin{array}{r}
-4.9217401366579165009139555520750\;\;
\\70060721450958436559876\times 10^8
\end{array}
 & 10^{-47}

\end{array}\nonumber
\eeq

We found that our result \eq{res} reproduces perfectly the first line in the table within the numerical error $10^{-61}$
which leaves no room for doubt in the validity of our result.

\section{Summary}
In this letter we have applied the Quantum Spectral Curve method \cite{PmuPRL,PmuLong} to the calculation of the NNLO correction to the BFKL eigenvalue.
We check our result numerically with high precision using the algorithm developed in \cite{Gromov:2015wca}
and gave numerical predictions for a few next orders.
We also developed a general efficient analytic method suitable for systematic perturbative solution of QSC.

There are numerous packages such as \cite{packages} available for the evaluation of the nested harmonic sums. Yet to simplify future applications of our results we attached a small Mathematica notebook \cite{NB} which allows to evaluate our result numerically and also generate expansions about singularities and at infinity.

We hope that our findings could shed some light on the QCD counterpart of our result
and resolve some mysteries shrouding the BFKL physics.

  \begin{acknowledgments}
\label{sec:acknowledgments}
We thank M.~Alfimov, B.~Basso, S.~Leurent and especially S.~Caron-Huot and V.~Kazakov for discussions. We are also very grateful to B.~Basso for verifying our result against the large $\Delta$ prediction from \cite{Basso:2014pla}.
The research leading to these results has received funding from the
People Programme (Marie Curie Actions) of the European Union's Seventh
Framework Programme FP7/2007-2013/ under REA Grant Agreement No 317089.
We wish to thank
SFTC for support from Consolidated
grant number ST/J002798/1.
N.G. would like to thank FAPESP grant 2011/11973-4 for funding his visit to ICTP-SAIFR during January 2015 where part of this work was done.

\end{acknowledgments}

\end{document}